\documentclass[11pt]{article}
\usepackage[margin=1.5in]{geometry}

\usepackage{mdwlist}
\usepackage{enumerate}

\usepackage{amsmath,amssymb}
\usepackage{graphics, subfigure, float}
\usepackage{fp, calc}
\usepackage{hyperref}
\usepackage{url}

\usepackage[T1]{fontenc} 
\usepackage{fourier}
\usepackage{bm}

\usepackage{amscd,amsthm}

\usepackage{pst-all}
\usepackage{pstricks-add}
\usepackage{pst-func}
\newpsobject{showgrid}{psgrid}{subgriddiv=1,griddots=10,gridlabels=6pt}
\usepackage{verbatim, comment}
\usepackage{datetime}

\newtheoremstyle{theorem}{1em}{1em}{\slshape}{0pt}{\bfseries}{.}{ }{}
\theoremstyle{theorem}
\newtheorem{theorem}{Theorem}
\newtheorem{question}{Question}
\newtheorem{conjecture}{Conjecture}
\newtheorem*{theorem*}{Theorem}
\newtheorem{corollary}[theorem]{Corollary}

\newtheorem{lemma}[theorem]{Lemma}

\newtheorem*{claim*}{Claim}

\theoremstyle{remark}

\newtheorem*{remark*}{Remark}

\newtheoremstyle{example}{1em}{1em}{}{0pt}{\bfseries}{.}{ }{}

\providecommand{\setR}{\mathbb{R}}

\newcommand{\tr}{\textrm{tr}}

\newcommand{\supp}{\textrm{supp}}

\newcommand{\E}{\mathop{\mathbb{E}}}

\newcommand{\eps}{\varepsilon}

\psset{linewidth=1pt, arrowsize=6pt}

\usepackage{calrsfs} 
\DeclareMathAlphabet{\pazocal}{OMS}{zplm}{m}{n}
\newcommand{\I}{\pazocal{I}}
\newcommand{\qedclaim}{$\diamond$}

\usepackage[displaymath,textmath,graphics, subfigure, floats]{preview} 
\PreviewEnvironment{center} 

\makeatother

\title{Linear Size Sparsifier and the Geometry of the Operator Norm Ball}
\date{}
\author{Victor Reis \thanks{University of Washington, Seattle. Email: {\tt voreis@uw.edu}.} \and Thomas Rothvoss  \thanks{University of Washington, Seattle. Email: {\tt rothvoss@uw.edu}. 
Supported by NSF CAREER grant 1651861 and a David \& Lucile Packard Foundation Fellowship.}}

\begin{document}
\maketitle
\begin{abstract}
The Matrix Spencer Conjecture asks whether given $n$ symmetric matrices in $\setR^{n \times n}$
with eigenvalues in $[-1,1]$ one can always find signs so that their signed sum has singular values
bounded by $O(\sqrt{n})$. The standard approach in discrepancy requires proving that the
convex body of all good fractional signings is large enough. However, this question has remained
wide open due to the lack of tools to certify measure lower bounds for rather small non-polyhedral convex sets.

A seminal result by Batson, Spielman and Srivastava from 2008 shows that any undirected graph 
admits a linear size spectral sparsifier. Again, one can define a convex body of all good fractional signings. We can indeed prove that this body is close to most of the Gaussian measure. 
This implies that a discrepancy algorithm by the second author can be used to sample a 
linear size sparsifer. In contrast to previous methods, we require only a logarithmic number of sampling
phases. 
\end{abstract}

\section{Introduction}

\emph{Discrepancy theory} is a subfield of combinatorics with several applications to theoretical computer science, see for example the books ~\cite{GeometricDiscrepancy-Matousek-1999, DiscrepancyMethod-Chazelle-2000}. 
In the classical setting one is given a family of sets $\pazocal{S} = \{ S_1,\ldots,S_m\}$ with $S_i \subseteq \{1,\ldots,n\}$ and the goal is to find a \emph{coloring} $\chi : [n] \to \{ -1,+1\}$ so that the maximum imbalance
$\max_{S \in \pazocal{S}} |\sum_{j \in S}\chi(j)|$ is minimized. This minimum value is called the \emph{discrepancy} of the family, denoted by $\textrm{disc}(\pazocal{S})$. A seminal result of Spencer~\cite{SixStandardDeviationsSuffice-Spencer1985} says that for any set family one has $\textrm{disc}(\pazocal{S}) \leq O(\sqrt{n \log (2m/n)})$, assuming that $m \geq n$. It is instructive to observe that for $m=n$, Spencer's result gives the bound of $O(\sqrt{n})$, while a uniform random coloring will have a discrepancy of $O(\sqrt{n\log(n)})$.
Moreover, one can show that for some set systems, only an exponentially small fraction of all colorings will indeed have a discrepancy of  $O(\sqrt{n})$. This demonstrates that in fact, Spencer's result provides the existence of a rather rare object. 

The cleanest approach to prove Spencer's result is due to Giannopoulos~\cite{Giannopoulos1997}, which we sketch for $m=n$: Consider the set $K = \{ \bm{x} \in \setR^n : |\sum_{j \in S_i} x_j| \leq \sqrt{n} \; \forall i \in [n] \} = \bigcap_{i \in [n]} Q_i$, a symmetric convex body which denotes the set of good-enough fractional colorings. Here $Q_i$ is the
strip of colorings that are good for set $S_i$. The Lemma of Sidak-Khatri~\cite{KhatriCorrelationInequality67,SidaksLemma67} allows us 
to lower bound the \emph{Gaussian measure} of $K$ as $\gamma_n(K) \geq \prod_{i=1}^n \gamma_n(Q_i) \geq e^{-cn}$ for some constant $c>0$ using that each strip $Q_i$ has a constant width. This rather weak bound on the measure is sufficient to use a \emph{pigeonhole principle} argument and conclude that  $c'K$ must contain a partial coloring $\bm{x} \in \{ -1,0,1\}^n$ with $|\textrm{supp}(\bm{x})| \geq \frac{n}{2}$.
Then one can color the elements in $\textrm{supp}(\bm{x})$ accordingly and repeat the argument for the remaining
uncolored elements. The overall  $O(\sqrt{n})$ bound follows from the fact that the discrepancy
of the partial colorings decreases geometrically as the number of elements in the set system decreases. 

While the pigeonhole principle based argument above is non-constructive in nature, Bansal~\cite{DiscrepancyMinimization-Bansal-FOCS2010} designed a polynomial time algorithm for finding the coloring guaranteed by Spencer's Theorem. Here, \cite{DiscrepancyMinimization-Bansal-FOCS2010}
exploits that it suffices to obtain a good enough \emph{fractional} partial coloring $\bm{x} \in [-1,1]^n$ with
a constant fraction of entries in $\{ -1,1\}$ to make the argument work. Later, Lovett and Meka~\cite{DiscrepancyMinimization-LovettMekaFOCS12} found a Brownian motion-type algorithm 
that --- despite being a lot simpler --- works for more general polyhedral settings. Finally, the random projection algorithm of Rothvoss~\cite{ConstructiveDiscrepancy-Rothvoss-FOCS2014} works for arbitrary symmetric
convex bodies that satisfy the measure lower bound. Another remarkable result is due to Bansal, Dadush, Garg and Lovett~\cite{GramSchmidtWalk-BansalDGL-STOC18}: for any symmetric body $K$ with $\gamma_n(K) \geq \frac{1}{2}$ and any vectors $\bm{v}_1,\ldots,\bm{v}_m \in \setR^n$ of 
length $\|\bm{v}_i\|_2 \leq 1$, one can find signs $\bm{x} \in \{ -1,1\}^m$ in randomized polynomial time so 
that $\sum_{i=1}^m x_i\bm{v}_i \in O(1) \cdot K$. This was known before by a non-constructive convex geometric argument
due to Banaszczyk~\cite{BalancingVectors-Banaszczyk98}.

There are two possible strengthenings of Spencer's Theorem that are both open at the time of this writing: suppose
that the set system is \emph{sparse} in the sense that every element is in at most $t$ sets. It is known that $\textrm{disc}(\pazocal{S}) \leq 2t$~\cite{IntegerMakingTheorems-BeckFiala81}
as well as $\textrm{disc}(\pazocal{S}) \leq O(\sqrt{t\log(n)})$~\cite{BalancingVectors-Banaszczyk98,GramSchmidtWalk-BansalDGL-STOC18}, while the Beck-Fiala Conjecture suggests that $\textrm{disc}(\pazocal{S}) \leq O(\sqrt{t})$ is
the right bound. For the second generalization --- the one that we are following in this paper --- it is helpful
to define $\bm{A}_i$ as the $m \times m$ diagonal matrix with $(j, j)$ entry $1$ if $i \in S_j$ and $0$ otherwise. If $\| \cdot \|_{\textrm{op}}$
denotes the maximum singular value of a matrix, then Spencer's result can be interpreted as the
existence of a coloring $\bm{x} \in \{ -1,1\}^n$ so that $\| \sum_{i=1}^n x_i\bm{A}_i\|_{\textrm{op}} \leq O(\sqrt{n \log(2m/n)})$. A conjecture raised by Meka\footnote{See the blog post \url{https://windowsontheory.org/2014/02/07/discrepancy-and-beating-the-union-bound/}} is whether for $m=n$, this bound is also possible for arbitrary symmetric matrices $\bm{A}_{1},\ldots,\bm{A}_n \in \setR^{n \times n}$ that satisfy $\|\bm{A}_i\|_{\textrm{op}} \leq 1$. One can prove using \emph{matrix concentration inequalities} that a random coloring $\bm{x}$ will lead to $\|\sum_{i=1}^n x_i\bm{A}_i\|_{\textrm{op}} \leq O(\sqrt{n\log(n)})$, and the same bound can also be achieved deterministically using a matrix multiplicative weight update argument~\cite{MatrixBalancingZouziasICALP12}.
An excellent overview of matrix concentration can be found in the monograph of Tropp~\cite{MatrixConcentrationInequalities-Tropp15}.

To understand the difficulty of proving Meka's conjecture, assume $m=n$ and revisit the approach of 
Giannopoulos for Spencer's Theorem. We can again define a set 
\[
K := \Big\{ \bm{x} \in \setR^n : \Big\|\sum_{i=1}^n x_i\bm{A}_i\Big\|_{\textrm{op}} \leq \sqrt{n} \Big\} = \Big\{ \bm{x} \in \setR^n : \sum_{i=1}^n x_i \left<\bm{A}_i,\bm{y}\bm{y}^\top\right> \leq \sqrt{n} \;\; \forall \bm{y} \in \setR^m: \|\bm{y}\|_2=1\Big\}
\] 
of good enough fractional colorings. Since $\| \cdot \|_{\textrm{op}}$ is a norm, $K$ will indeed be symmetric and convex. 
It would hence suffice to prove that $\gamma_n(K) \geq 2^{-cn}$ for some constant $c>0$. However, it is open whether this
inequality holds. The issue is that $K$ is non-polyhedral and applying Sidak-Khatri's bound over infinitely\footnote{One can use an $\varepsilon$-net of $2^{\Theta(m)}$ many vectors $\bm{y}$ but the bound is still too weak.} many vectors $\bm{y}$
is way too inefficient. While matrix concentration inequalities are fantastic at proving that likely events are indeed likely, they seem to be unable to prove that unlikely events are not too unlikely. With a scaling argument, they can still be used to prove that $\gamma_n(K) \geq (\log(n))^{-cn}$ for some constant $c>0$, assuming $m=n$, though better bounds seem out of reach.

In terms of discrepancy in spectral settings, a different line of techniques has been arguably more successful. A beautiful and
influential paper by Batson, Spielman and Srivastava~\cite{TwiceRamanujanSparsifiers-BatsonSpielmanSrivastava-STOC09} proves that for any undirected graph on $n$ nodes one can take a weighted subgraph with just a \emph{linear} number of edges
that approximates every cut within a constant factor. Translated into linear algebra terms, \cite{TwiceRamanujanSparsifiers-BatsonSpielmanSrivastava-STOC09} show that given any vectors $\bm{v}_1,\ldots,\bm{v}_m \in \setR^n$ that are in 
\emph{isotropic position}, i.e. $\sum_{i=1}^m \bm{v}_i\bm{v}_i^\top = \bm{I}_n$, one can find weights $\bm{s} \in \setR_{\geq 0}^m$ with
$|\textrm{supp}(\bm{s})| \leq O(n/\varepsilon^2)$ so that $(1-\varepsilon) \cdot \bm{I}_n \preceq \sum_{i=1}^m s_i\bm{v}_i\bm{v}_i^\top \preceq (1+\varepsilon) \cdot \bm{I}_n$, and indeed Lee and Sun showed this can be done in nearly linear time~\cite{YinTatSparsifier}. In a more recent ce{}lebrated paper, Marcus, Spielman and Srivastava~\cite{MarcusSpielmanSrivastava-SolutionOfKadisonSinger-AnnalsOfMath2015} resolved the \emph{Kadison-Singer Conjecture}, a problem that has appeared independently in different forms in many areas of mathematics. In a simple-to-state version,
their result says that for any vectors $\bm{v}_1,\ldots,\bm{v}_m \in \setR^n$ with $\sum_{i=1}^m \bm{v}_i\bm{v}_i^\top = \bm{I}_n$ and $\|\bm{v}_i\|_2 \leq \varepsilon$ for all $i \in [m]$, 
there are signs $\bm{x} \in \{ -1,1\}^m$ so that $\|\sum_{i=1}^m x_i\bm{v}_i\bm{v}_i^\top \|_{\textrm{op}} \leq O(\varepsilon)$.
On a very high level view, both methods of \cite{TwiceRamanujanSparsifiers-BatsonSpielmanSrivastava-STOC09} and \cite{MarcusSpielmanSrivastava-SolutionOfKadisonSinger-AnnalsOfMath2015} control a carefully
chosen potential function, though we note there is still no known polynomial time algorithm for the latter. 

The goal of this paper will be to connect the classical discrepancy theory and the spectral discrepancy theory of 
\cite{TwiceRamanujanSparsifiers-BatsonSpielmanSrivastava-STOC09, MarcusSpielmanSrivastava-SolutionOfKadisonSinger-AnnalsOfMath2015} and develop arguments that prove largeness of non-polyhedral bodies. We remark that we made no attempt at
optimizing constants but rather prefer to keep the exposition simple. 

\paragraph{Notation.}
For a (not necessarily symmetric) matrix $\bm{M} \in \setR^{n \times n}$ the operator norm can be formally
defined as $\|\bm{M}\|_{\textrm{op}} := \max\{ \|\bm{M}\bm{x}\|_2 : \bm{x} \in \setR^n\textrm{ with }\|\bm{x}\|_2 = 1\}$.
For a symmetric matrix $\bm{A} \in \setR^{n \times n}$ with eigendecomposition $\bm{A} = \sum_{i=1}^n \lambda_i\bm{v}_i\bm{v}_i^\top$, we write $|\bm{A}| := \sum_{i=1}^n |\lambda_i| \bm{v}_i\bm{v}_i^\top$ as the matrix where all eigenvalues have been replaced by their absolute values. In this notation, $\|\bm{A}\|_{\textrm{op}} := \max\{ |\lambda_i| : i \in [n] \}$ is the maximum singular value. We abbreviate $B_2^n := \{ \bm{x} \in \setR^n \mid \|\bm{x}\|_2 \le 1\}$ and $S^{n-1} := \{ \bm{x} \in \setR^n \mid \|\bm{x}\|_2=1\}$. 
Given symmetric matrices $\bm{A}, \bm{B} \in \setR^{n \times n}$, we write $\bm{A} \preceq \bm{B}$ if $\bm{x}^\top \bm{A} \bm{x} \le \bm{x}^\top \bm{B} \bm {x}$ for all $\bm{x} \in \setR^n$.

A \emph{convex body} is a closed convex set $K \subset \setR^n$ with nonempty interior. We denote $d(\bm{y}, K) := \min_{\bm{x} \in K} \|\bm{x} - \bm{y}\|_2$ as the \emph{distance} from $\bm{y}$ to $K$. Let $K_{\delta} = \{\bm{x} \in \setR^n \mid d(\bm{x}, K) \le \delta\}$ be the set of points that have distance at most $\delta$ to $K$ (in particular, $K \subseteq K_\delta$). The \emph{Minkowski sum} of sets $A$ and $B$ is defined as $A+B:= \{\bm{a}+\bm{b} \mid \bm{a} \in A, \bm{b} \in B\}$. A \emph{halfspace} is a set of the form $H := \{\bm{x} \in \setR^n | \left<\bm{v},\bm{x}\right> \le \lambda\}$ for some $\bm{v} \in \setR^n$ and $\lambda \in \setR$. The \emph{Gaussian measure} of $K$ is defined as $\gamma_n(K) := \Pr_{\bm{y} \sim N(\bm{0},\bm{I}_n)}[\bm{y} \in K]$. 
Here $N(\bm{0},\bm{I}_n)$ is the distribution of a standard Gaussian in $\setR^n$.

\subsection{Our contribution \label{sec:Contribution}}

A possible way to approach the setting of Batson, Spielman, Srivastava~\cite{TwiceRamanujanSparsifiers-BatsonSpielmanSrivastava-STOC09} from a classical discrepancy perspective is to take vectors $\bm{v}_1,\ldots,\bm{v}_m$ in isotropic position and
consider the body $K = \{ \bm{x} \in \setR^m \mid \|\sum_{i=1}^m x_i\bm{v}_i\bm{v}_i^\top\|_{\textrm{op}} \leq \sqrt{n/m} \}$. 
If we could prove that $\gamma_m(K) \geq 2^{-cm}$, then the algorithm of \cite{ConstructiveDiscrepancy-Rothvoss-FOCS2014} would be able to find a partial coloring.   
While we still do not know whether the inequality $\gamma_m(K) \geq 2^{-cm}$ holds, we can prove that a weaker condition that suffices for the algorithm of \cite{ConstructiveDiscrepancy-Rothvoss-FOCS2014}
is satisfied: 
\begin{theorem} \label{thm:MainResultGaussianExpansionOfK}
Let $\bm{A}_1,\ldots,\bm{A}_m \in \setR^{n \times n}$ be symmetric matrices with $\sum_{i=1}^{m} |\bm{A}_i| \preceq \bm{I}_n$ and select $\varepsilon \in (0, 1)$ so that $m = \frac{n}{\eps^2} \ge 100$. Then for any $0<\alpha < 1$, the set 
\[
 K := \Big\{ \bm{x} \in \setR^m \mid \Big\| \sum_{i=1}^m x_i \bm{A}_i \Big\|_{\textrm{op}} \leq \eps \Big\}
\]
satisfies $\gamma_m\Big(\frac{50}{\alpha} K + \alpha \sqrt{m} B_2^m\Big) \geq \frac{1}{2}$. That is, $\Pr_{\bm{y} \sim N(\bm{0}, \bm{I}_m)} \Big[d\Big(\bm{y}, \frac{50}{\alpha} K\Big) \le \alpha \sqrt{m}\Big] \ge \frac{1}{2}$.
\end{theorem}
Note that in particular the rank-1 matrices $\bm{A}_i = \bm{v}_i\bm{v}_i^T$ with $\sum_{i=1}^m \bm{v}_i\bm{v}_i^T \preceq \bm{I}_n$ satisfy the 
premise of Theorem~\ref{thm:MainResultGaussianExpansionOfK}.
A quantity that is often used in the convex geometry literature is the \emph{mean width} of a body $K$, which is defined as $ w(K) := \E_{\bm{a} \in S^{n-1}}[\max_{\bm{x} \in K} \left<\bm{a},\bm{x}\right>- \min_{\bm{x} \in K} \left<\bm{a},\bm{x}\right>]$. 
The above result implies the following: 
\begin{theorem}
A body $K$ as defined in Theorem~\ref{thm:MainResultGaussianExpansionOfK} has mean width $w(K) \geq \Omega(\sqrt{m})$.
\end{theorem}

A rather immediate consequence of this insight is that the following sampling algorithm will
work with very high probability:  
\begin{center}
\psframebox{
\begin{minipage}{13.2cm}
{\sc Spectral Sparsification Algorithm} \vspace{1mm} \hrule \vspace{1mm}
\begin{itemize*}
\item {\bf Input:} PSD matrices $\bm{A}_1,\ldots,\bm{A}_m \in \setR^{n \times n}$ with $\sum_{i=1}^m \bm{A}_i = \bm{I}_n$ and  $\varepsilon>0$ \vspace{0mm}
\item {\bf Output:} $\bm{s} \in \setR_{\geq 0}^m$ with $\textrm{supp}(\bm{s}) \leq \frac{n}{\varepsilon^2}$ and $(1-O(\varepsilon)) \bm{I}_n \preceq \sum_{i=1}^m s_i\bm{A}_i \preceq (1+O(\varepsilon)) \bm{I}_n$ \vspace{1mm}\hrule \vspace{-3mm}
\end{itemize*} 
\begin{enumerate*}
\item[(1)] Set $s_i := 1$ for $i \in [m]$ 
\item[(2)] WHILE $|\textrm{supp}(\bm{s})| > \frac{n}{\varepsilon^2}$ DO
  \begin{enumerate*}
  \item[(3)] Let 
$
 K := \{ \bm{x} \in \setR^{\textrm{supp}(\bm{s})} : \|\sum_{i \in \textrm{supp}(\bm{s})} x_is_i\bm{A}_i\|_{\textrm{op}} \leq 1000\tilde{\varepsilon} \}
$ with $|\textrm{supp}(\bm{s})| = \frac{n}{\tilde{\varepsilon}^2}$.
  \item[(4)] Draw a Gaussian $\bm{y}^* \sim N(\bm{0},\bm{I}_{\textrm{supp}(\bm{s})})$.   
  \item[(5)] Compute $\bm{x}^* := \textrm{argmin}\{ \|\bm{x}-\bm{y}^*\|_2 : \bm{x} \in [-1,1]^{\textrm{supp}(s)} \cap K \}$.
  \item[(6)] If $\#(i:x_i^*=-1) < \#(i:x_i=1)$ then replace $\bm{x}^*$ by $-\bm{x}^*$. 
  \item[(7)] Update $s_i := s_i \cdot (1+x^*_i)$.
  \end{enumerate*}
\end{enumerate*}
\end{minipage}
}
\end{center}
In fact we will prove: 
\begin{theorem}
With probability at least  $1-2^{-\Omega(n)}$ a run of the {\sc Spectral Sparsification Algorithm} satisfies
all of the following properties: (a) the algorithm runs in polynomial time; 
(b) the while loop is iterated at most $O(\log m)$ times; (c) at the end one has $|\textrm{supp}(\bm{s})| \le \frac{n}{\eps^2}$
and  $(1-O(\eps)) \bm{I}_n \preceq \sum_{i=1}^m s_i\bm{A}_i \preceq (1+O(\eps)) \bm{I}_n$.
\end{theorem}
Note that our algorithm produces sparse vector $\bm{s}$ by iteratively finding
low discrepancy colorings. This technique has appeared before in the literature. For 
example for a set system with bounded \emph{VC dimension}, one can prove the existence of small \emph{$\varepsilon$-nets} in this manner. We refer to Chapter 4 of Chazelle's book~\cite{DiscrepancyMethod-Chazelle-2000} for details.

\section{Preliminaries}

In this section, we discuss several tools from probability and linear algebra that we will be
using in the proofs. 

\paragraph{Concentration.}

We need two concentration inequalities. 
For the first one, see~\cite{Handel2014ProbabilityIH}.
\begin{theorem}
If $F: \setR^m \to \setR$ is $1$-Lipschitz, then for $t \geq 0$ one has

\[
\Pr_{\bm{y} \sim N(\bm{0}, \bm{I}_m)} [F(\bm{y}) > \E[F(\bm{y})] + t] \le e^{-t^2/2}.
\]

\end{theorem}
For the proof of the following Corollary, see Appendix~\ref{sec:MissingProofsOfPrelim}.
\begin{corollary} \label{cor:ProbGaussianHasToBeTruncated}
For $m \ge 7$ we have 
\[
\Pr_{\bm{y} \sim N(\bm{0}, \bm{I}_m)} \Big[\|\bm{y}\|_2 > m\Big] \le 2^{-m}
\quad \textrm{and} \quad 
\E_{\bm{y} \sim N(\bm{0}, \bm{I}_m)}\Big[\|\bm{y}\|_2^2 \mid \|\bm{y}\| \le m\Big] \ge (1 - 2^{-m})\cdot m.
\]
\end{corollary}

We also need Azuma's inequality for Martingales with bounded increments, see~\cite{ProbabilisticMethod-AlonSpencer-Book2016}. 

\begin{theorem}[\textbf{Azuma's Inequality}] Let $0 = X_0, \dots, X_T$ be a Martingale with $|X_t - X_{t-1}| \le a$ for all $t = 1, \dots, T$. Then for any $\lambda \ge 0$ we have 
\[
\Pr[X_T > \lambda \sqrt{T} ] \le e^{-\lambda^2/2a^2}
\]
\end{theorem}

\paragraph{Gaussians.}

In order to increase the measure from $\frac{1}{2}$ to $1 - 2^{-\Omega(m)}$ we use the following key theorem, see~\cite{ProbabilityInBanachSpaces-LedouxTalagrand-Book2011}. 

\begin{theorem}[\textbf{Gaussian Isoperimetric Inequality}] Let $K \subset \setR^n$ be a measurable set and $H$ be a halfspace such that $\gamma_n(K) = \gamma_n(H)$. Then $\gamma_n(K_\delta) \ge \gamma_n(H_\delta)$ for all $\delta > 0$.
\end{theorem}

The following simple result is useful for dealing with dilations, see~\cite{TkoczThesis-2015}.

\begin{theorem} Let $K \subset \setR^n$ be a measurable set and $B$ be a closed Euclidean ball such that $\gamma_n(K) = \gamma_n(B)$. Then $\gamma_n(tK) \ge \gamma_n(tB)$ for all $t \in [0, 1]$.
\end{theorem}


For (not neccesarily symmetric) matrices $\bm{A},\bm{B} \in \setR^{n \times n}$ we  define the \emph{Frobenius inner product} $\left<\bm{A},\bm{B}\right>_F := \tr[\bm{A}^\top\bm{B}] = \sum_{i=1}^n\sum_{j=1}^n A_{ij}B_{ij}$ and the corresponding 
\emph{Frobenius norm} $\|\bm{A}\|_F := \sqrt{\left<\bm{A},\bm{A}\right>_F} = (\sum_{i=1}^n\sum_{j=1}^n A_{ij}^2)^{1/2}$. 
Generalizing earlier notation, for a PSD matrix $\bm{X} \in \setR^{m \times m}$, we define $N(\bm{0},\bm{X})$ as the distribution
of a centered Gaussian with covariance matrix $\bm{X}$. Note that there is a canonical way to generate such a distribution: 
let $X_{ij} = \left<\bm{v}_i,\bm{v}_j\right>$ be the factorization of that matrix for some vectors $\bm{v}_i \in \setR^r$. Then draw a standard Gaussian
$\bm{y} \sim N(\bm{0},\bm{I}_r)$, so that $(\left<\bm{g},\bm{v}_1\right>,\ldots,\left<\bm{g},\bm{v}_m\right>) \sim N(\bm{0},\bm{X})$. In particular we will be interested in drawing
a standard Gaussian restricted to a subspace $H \subseteq \setR^m$. The distribution of such a Gaussian is exactly $N(\bm{0},\bm{X})$
where $\bm{X} = \sum_{i=1}^{\dim(H)} \bm{u}_i\bm{u}_i^\top$ and $\bm{u}_1,\ldots,\bm{u}_{\dim(H)}$ is an orthonormal basis of $H$. The following properties are well known: 
\begin{lemma} \label{lem:PropertiesGaussianFromSubspace}
Let $H \subseteq \setR^m$ be a subspace and let $N(\bm{0},\bm{X})$ be the distribution of a standard Gaussian
restricted to that subspace. Then for $\bm{y} \sim N(\bm{0},\bm{X})$ one has
$(i)$ $\bm{y} \in H$ always; (ii) $\E[\|\bm{y}\|_2^2]=\textrm{Tr}[\bm{X}]=\dim(H)$; (iii) $\E[y_i^2] \leq 1$ for all $i \in [m]$; (iv)
$\textrm{Var}[\left<\bm{y},\bm{b}\right>] = \E[\left<\bm{y},\bm{b}\right>^2]\leq \|\bm{b}\|_2^2$ for all $\bm{b} \in \setR^m$; (v) for any matrices $\bm{W}^1,\ldots,\bm{W}^m \in \setR^{n \times n}$ one has $\E[\|\sum_{i=1}^m y_i\bm{W}^i\|_F^2] \leq \sum_{i=1}^m \|\bm{W}^i\|_F^2$.
\end{lemma}
The only property that is non-standard is $(v)$. But note that we can use $(iv)$ to justify that for each entry $(k,\ell)$
of the matrices one has $\E[(\sum_{i=1}^m y_iW_{k\ell}^i)^2] \leq \sum_{i=1}^m (W_{k\ell}^i)^2$; the claim then follows
by linearity of expectation and summing over all entries $(k,\ell) \in [n]^2$.

\paragraph{Linear Algebra.}

For the analysis, we need an estimate on the trace of the product of symmetric matrices.
The proof takes some care due to the non-commuting matrices.
To get some intuition, consider the case when $\bm{A}_1,\bm{A}_2,\bm{B}$ are all diagonal matrices.
In this case one can write $\bm{A}_1\bm{B} = \textrm{diag}(\bm{a}_1)$ and $\bm{A}_2\bm{B} = \textrm{diag}(\bm{a}_2)$ for some vectors $\bm{a}_1,\bm{a}_2 \in \setR^n$ and the 
inequality simplifies to $\tr[\bm{A}_1\bm{B}\bm{A}_2\bm{B}] = \left<\bm{a}_1,\bm{a}_2\right> \leq \|\bm{a}\|_1 \cdot \|\bm{a}_2\|_1= \tr[\bm{A}_1|\bm{B}|] \cdot \tr[\bm{A}_2|\bm{B}|]$
which is obviously true. Note that in the setting of~\cite{TwiceRamanujanSparsifiers-BatsonSpielmanSrivastava-STOC09} we would apply Lemma~\ref{lem:TraceOfProductOfMatrices} with $\textrm{rank}(\bm{B}) = 2$,  in which case the inequality can be tight up to constant factors. But in a different application 
with higher-rank matrices one could imagine a Cauchy-Schwarz or H\"older-type inequality yielding
improved bounds.
\begin{lemma} \label{lem:TraceOfProductOfMatrices}
Let $\bm{A}_1, \bm{A}_2, \bm{B} \in \setR^{n \times n}$ be symmetric matrices with $\bm{A}_1, \bm{A}_2 \succeq 0$. Then 
$$\tr[\bm{A}_1 \bm{BA}_2 \bm{B}] \le \tr[\bm{A}_1|\bm{B}|] \cdot \tr[\bm{A}_2|\bm{B}|].$$
\end{lemma}

\begin{proof}
Write the spectral decomposition $\bm{B} = \sum_{i \in [n]} \lambda_i \bm{v}_i \bm{v}_i^\top$. Then
  \begin{align*}
  \tr[\bm{A}_1 \bm{BA}_2 \bm{B}] &= \sum_{i, j \in [n]} \lambda_i \lambda_j \cdot \big(\bm{v}_i^\top \bm{A}_1 \bm{v}_j\big)\big(\bm{v}_i^\top  \bm{A}_2 \bm{v}_j\big) 
  \\ &\le \sum_{i, j \in [n]} |\lambda_i| \cdot |\lambda_j| \cdot \big\|\bm{A}_1^{1/2} \bm{v}_i\big\|_2 \cdot \big\|\bm{A}_2^{1/2} \bm{v}_i\big\|_2 \cdot \big\|\bm{A}_1^{1/2} \bm{v}_j\big\| \cdot \big\|\bm{A}_2^{1/2} \bm{v}_j\big\|_2 
  \\ &\le \sum_{i, j \in [n]}|\lambda_i| \cdot |\lambda_j| \cdot \frac{1}{2} \Big(\|\bm{A}_1^{1/2} \bm{v}_i\|^2_2 \cdot \|\bm{A}_2^{1/2} \bm{v}_j\|^2 + \|\bm{A}_1^{1/2} \bm{v}_j\|^2_2 \cdot \|\bm{A}_2^{1/2} \bm{v}_i\|^2_2 \Big) 
  \\ &= \sum_{i, j \in [n]}|\lambda_i| |\lambda_j| \cdot \frac{1}{2} \Big( \big(\bm{v}_i^\top \bm{A}_1 \bm{v}_i\big) \cdot \big(\bm{v}_j^\top \bm{A}_2 \bm{v}_j\big) + \big(\bm{v}_i^\top \bm{A}_2 \bm{v}_i\big) \cdot \big(\bm{v}_j^\top \bm{A}_1 \bm{v}_j\big) \Big) 
  \\ &= \frac{1}{2} \Big( \tr[\bm{A}_1|\bm{B}|]\cdot \tr[\bm{A}_2|\bm{B}|] + \tr[\bm{A}_2|\bm{B}|]\cdot \tr[\bm{A}_1|\bm{B}|] \Big) 
  \\ & = \tr[\bm{A}_1|\bm{B}|] \cdot \tr[\bm{A}_2|\bm{B}|], \end{align*} 
 where the first inequality is Cauchy-Schwarz and the second is AM-GM.
\end{proof}

We also need a Taylor approximation for the trace of the inverse of a matrix.
Again, it takes some care to handle the non-commutativity:

\begin{lemma} \label{lem:MatrixTaylorApprox}
Let $\bm{A},\bm{B} \in \setR^{n \times n}$ be symmetric matrices with $\bm{A} \succ 0$ and $\|\delta \bm{A}^{-1} \bm{B}\|_{\textrm{op}} \leq \frac{1}{2}$.
Then there is a value $c := c(\bm{A},\bm{B},\delta) \in [-2,2]$ so that
\[
 \tr[(\bm{A}-\delta \bm{B})^{-1}] = \tr[\bm{A}^{-1}] + \delta \tr[\bm{A}^{-1}\bm{B}\bm{A}^{-1}] + c\delta^2 \tr[\bm{A}^{-1}\bm{B}\bm{A}^{-1}\bm{B}\bm{A}^{-1}].
\]
\end{lemma}

\begin{proof}
We abbreviate $\bm{M} := \delta \bm{A}^{-1}\bm{B}$. As $\|\bm{M}\|_{\textrm{op}} \leq \frac{1}{2}$, the matrix $\bm{I}_n-\bm{M}$
is non-singular and by direct computation one can verify that its inverse is given by  $(\bm{I}_n - \bm{M})^{-1} = \bm{I}_n + (\bm{I}_n - \bm{M})^{-1} \bm{M}$. Using this formula twice at $(*)$, we obtain
\begin{align*}
(\bm{A}-\delta \bm{B})^{-1} &= \big(\bm{A} (\bm{I}_n - \delta \bm{A}^{-1}\bm{B})\big)^{-1} \\
&= (\bm{I}_n - \bm{M})^{-1}\bm{A}^{-1} \\ 
&\stackrel{(*)}{=} \bm{A}^{-1} + (\bm{I}_n-\bm{M})^{-1}\bm{M}\bm{A}^{-1} \\
&\stackrel{(*)}{=} \bm{A}^{-1} + \bm{M}\bm{A}^{-1} + (\bm{I}_n - \bm{M})^{-1}\bm{M}\bm{M}\bm{A}^{-1} \\
&= \bm{A}^{-1} + \delta\bm{A}^{-1} \bm{B} \bm{A}^{-1} + \delta^2 (\bm{I}_n - \bm{M})^{-1} \bm{A}^{-1} \bm{B} \bm{A}^{-1} \bm{B} \bm{A}^{-1}.
\end{align*}
Taking the trace on both sides gives 
\[
\tr[ (\bm{A} - \delta \bm{B})^{-1}] = \tr[ \bm{A}^{-1}] + \delta \tr[ \bm{A}^{-1} \bm{B} \bm{A}^{-1}] + \delta^2 \tr[(\bm{I}_n - \bm{M})^{-1} \bm{A}^{-1} \bm{B} \bm{A}^{-1} \bm{B} \bm{A}^{-1}].
\]
Since $\bm{A}^{-1} \succ 0$, we have $\bm{A}^{-1} \bm{B} \bm{A}^{-1} \bm{B} \bm{A}^{-1} \succeq 0$, hence we can bound the absolute value of the last term as
\[
|\tr[(\bm{I}_n - \bm{M})^{-1} \bm{A}^{-1} \bm{B} \bm{A}^{-1} \bm{B} \bm{A}^{-1}]| \le \|(\bm{I}_n - \bm{M})^{-1} \|_{\textrm{op}} \cdot \tr[\bm{A}^{-1} \bm{B} \bm{A}^{-1} \bm{B} \bm{A}^{-1}].
\]
Finally, note that 
\[
\big\|(\bm{I}_n - \bm{M})^{-1}\big\|_{\textrm{op}} = \Big\|\sum_{k=0}^\infty \bm{M}^k\Big\|_{\textrm{op}} \le \sum_{k=0}^\infty \|\bm{M}\|^k_{\textrm{op}} \le 2.
\]
\end{proof}

\section{Main technical result}

We now show our main result, Theorem~\ref{thm:MainResultGaussianExpansionOfK}. Fix symmetric matrices $\bm{A}_1,\ldots,\bm{A}_m \in \setR^{n \times n}$ with $\sum_{i=1}^m |\bm{A}_i| \preceq \bm{I}_n$ and set $\varepsilon>0$ so that $m = \frac{n}{\varepsilon^2}$. Let $K$ be the body as defined in Theorem~\ref{thm:MainResultGaussianExpansionOfK} and fix a parameter $\alpha>0$. Ideally, the goal would be to prove that a random Gaussian from $N(\bm{0},\bm{I}_m)$ is on average close to $K$.
Instead, we prove that there is a random variable $\bm{x}$ that is close to a Gaussian and ends up in $K$
with high probability.
 The strategy is to generate such a near-Gaussian random variable $\bm{x}$ by performing a \emph{Brownian motion} that adds up independent Gaussians $\bm{y}^{(t)}$ with a tiny step size $\delta$. The key ingredient is that in each iteration $t$ we walk inside a subspace of dimension at least $(1-\alpha^2)m$, meaning that we draw $\bm{y}^{(t)} \sim N(\bm{0},\bm{X}^{(t)})$ with $\textrm{tr}[\bm{X}^{(t)}] \geq (1-\alpha^2)m$. This can be understood as blocking the movement in $\alpha^2m$ dimensions that are ``dangerous''.
Then the expected Euclidean distance of the outcome $\bm{x} = \delta \sum_{t=1}^{1/\delta^2} \bm{y}^{(t)}$ to an unrestricted Gaussian is at most $\alpha \sqrt{m}$. It remains to argue that the subspace can be chosen so that at the end of the Brownian motion,
 $\bm{x}$ ends up in $K$. 
For this sake we define a \emph{potential function}
\[
 \bm{A}_{C,D}(\bm{x}) :=  (C+D\|\bm{x}\|_2^2) \cdot \bm{I}_n - \sum_{i=1}^m x_i \bm{A}_i \quad \textrm{and} \quad  \Phi_{C,D}(\bm{x}) := \tr\left[\bm{A}_{C,D}(\bm{x})^{-1}\right] 
\]
We initialize the random walk with $\bm{x} := \bm{0}$ so that $\bm{A}_{C,D}(\bm{x}) \succ 0$. 
If the update steps are small and we keep the potential function $\Phi_{C,D}(\bm{x})$ bounded, 
we can infer that  $\sum_{i=1}^m x_i\bm{A}_i \preceq (C+D\|\bm{x}\|_2^2) \cdot \bm{I}_n$ at any given
time. 
More precisely we show that, for a particular choice of parameters $C,D>0$ (later we will choose $C = \Theta(\frac{\varepsilon}{\alpha})$ and $D = \Theta(\frac{\varepsilon}{\alpha m})$), an update of $\bm{x}' = \bm{x} + \delta \bm{y}^{(t)}$ in expectation does not
increase the value of the potential function --- assuming that the current value of the potential function is small enough
and $\bm{y}^{(t)}$ is taken from the aforementioned subspace. 

In order to get some more intuition behind the potential function, let us discuss why a potential function  $\tilde{\Phi}_C(\bm{x}) = \tr[ \tilde{\bm{A}}_{C}(\bm{x})^{-1}] = \tr[(C \cdot \bm{I}_n - \sum_{i=1}^m x_i \bm{A}_i)^{-1}]$ with a fixed barrier term would be problematic: if we update $\bm{x}'  = \bm{x} + \delta \bm{y}^{(t)}$ so that $\E[\bm{x}']=\bm{x}$, 
then by \emph{strict convexity} of the function $z \mapsto \frac{1}{1-z}$, one has in general $\E[\tilde{\Phi}_C(\bm{x}')] > \tilde{\Phi}_C(\bm{x})$. This is where the additional ``variance term'' $\|\bm{x}\|_2^2$ comes into play. We know that $\|\bm{x}'\|_2^2 = \|\bm{x}\|_2^2 + \delta^2 \|\bm{y}^{(t)}\|_2^2 $ (assuming we pick the update direction orthogonal to $\bm{x}$). Then in every update step, the barrier is shifted a bit. It remains to show that the decrease from the barrier shift can compensate for the increase due to strict convexity.

There is the technical issue that the potential function goes up to $\infty$
as the minimal eigenvalue of $\bm{A}_{C,D}(\bm{x})$ approaches $0$. We solve this problem by
defining another distribution $N_{\leq m}(\bm{0},\bm{X})$ that draws $\bm{y} \sim N(\bm{0},\bm{X})$, but if $\|\bm{y}\|_2 > m$, then $\bm{y}$ is replaced with $\bm{0}$. Recall that by Corollary 5 one has $\Pr_{\bm{y} \sim N(\bm{0},\bm{X})}[\|\bm{y}\|_2>m] \leq 2^{-m}$ for any $\bm{X} \preceq \bm{I}_m$. A second problem is that keeping the potential function
low in expectation is not sufficient --- if the potential function ever crosses a certain threshold, the analysis stops working. 
However, a single step in the Brownian motion can be analyzed as follows:

\begin{lemma} \label{lem:PotentialFunctionUpdateInOneIteration}
Fix $0<\alpha<1$ and $m \ge \max\Big\{100, \frac{10}{\alpha^2}\Big\}$.
Let $\bm{x} \in \setR^m$ and suppose $\bm{A}_{C,D}(\bm{x}) \succ 0$, $\Phi_{C,D}(\bm{x}) \leq \frac{Dm^2\alpha^2}{10}$ as well as $0< \delta \leq \frac{1}{5Dm^5}$.
Define  $S(\bm{y})$ as the unique value for which 
\[
\Phi_{C + \delta^2 S(\bm{y}),D}(\bm{x}+\delta \bm{y}) = \Phi_{C,D}(\bm{x}). 
\]
Then there is a covariance matrix $\bm{X} \in \setR^{m \times m}$ with $0 \preceq \bm{X} \preceq \bm{I}_m$ and $\textrm{Tr}[\bm{X}] \geq (1-\alpha^2) \cdot m$ so that $\E_{\bm{y} \sim N_{\leq m}(\bm{0},\bm{X})}[S(\bm{y})] \le 0$
while always $|S(\bm{y})| \leq 4Dm^4$. Further, $\bm{A}_{C + \delta^2 S(\bm{y}), D} (\bm{x} + \delta \bm{y}) \succ 0$.
\end{lemma}

We postpone the proof of this lemma to Section 4. First, we show how we can use it to obtain the main theorem:

\begin{proof}[Proof of Theorem~\ref{thm:MainResultGaussianExpansionOfK}] 
Let $\bm{A}_1,\ldots,\bm{A}_m \in \setR^{n \times n}$ be symmetric matrices with $\sum_{i=1}^{m} |\bm{A}_i| \preceq \bm{I}_n$ so that $m = \frac{n}{\eps^2} \ge 100$. 
Fix a parameter $0<\alpha<1$ and keep in mind that the goal is to prove that $\gamma_m\Big(\frac{50}{\alpha} K + \alpha \sqrt{m} B_2^m\Big) \geq \frac{1}{2}$.
Note that the potential function $\Phi_{C,D}(\bm{x})$ is \emph{one-sided} in the sense that it only controls the \emph{maximum} eigenvalue of $\sum_{i=1}^m x_i\bm{A}_i$. For this sake we abbreviate
\[
\tilde{\bm{A}}_i := \begin{pmatrix} \bm{A}_i & \bm{0} \\ \bm{0} & -\bm{A}_i \end{pmatrix} \in \setR^{2n \times 2n}
\]
Note that this allows us to rewrite $K = \Big\{ \bm{x} \in \setR^m \mid \sum_{i=1}^m x_i\tilde{\bm{A}}_i \preceq \varepsilon \bm{I}_{2n}\Big\}$.
In wise foresight we choose $D := \frac{2 \eps}{\alpha m}$ and define $C$ so that $\frac{2n}{C} = \frac{Dm^2\alpha^2}{10}$ which results in $C = \frac{10\varepsilon}{\alpha}$. We define a small enough step size of $\delta := \frac{1}{5Dm^5}$ and choose  $T := \frac{1}{\delta^2}$ as the number of iterations. 

Note that by definition $\Phi_{C,D}(\bm{0}) = \frac{2n}{C} = \frac{Dm^2\alpha^2}{10}$.
Consider the following (hypothetical) algorithm: 
\begin{enumerate*}
\item[(1)] Set $\bm{x}^{(0)} := \bm{0}$
\item[(2)] For $t=1$ TO $T$ do
  \begin{enumerate*}
  \item[(3)] Apply Lemma~\ref{lem:PotentialFunctionUpdateInOneIteration} for $\bm{x}^{(t-1)}$ and let $\bm{X}^{(t)}$ be the obtained covariance matrix. 
  \item[(4)] Sample $\bm{y}^{(t)} \sim N(\bm{0},\bm{X}^{(t)})$ and $\bm{z}^{(t)} \sim N(\bm{0},\bm{I}_m-\bm{X}^{(t)})$.
  
  If $\|\bm{y}^{(t)}\|_2 \leq m$ then $(\bm{y}_{\leq m}^{(t)},\bm{y}_{>m}^{(t)}) := (\bm{y}^{(t)},\bm{0})$, otherwise $(\bm{y}_{\leq m}^{(t)},\bm{y}_{>m}^{(t)}) := (\bm{0},\bm{y}^{(t)})$.
  \item[(5)] Update $\bm{x}^{(t)} := \bm{x}^{(t-1)} + \delta \bm{y}_{ \leq m}^{(t)}$.
  \end{enumerate*}
\end{enumerate*}
At the end, let $\bm{Y} := \bm{x}^{(T)} = \delta \sum_{t=1}^T \bm{y}^{(t)}$ and $\bm{Z} := \delta \sum_{t=1}^T \bm{z}^{(t)}$. Note that $\bm{Y} + \bm{Z} \sim N(\bm{0},\bm{I}_m)$. 

{\bf Claim.} \emph{The following events all hold simultaneously with probability at least $\frac{1}{2}$:}
\begin{enumerate*}
\item[(a)] \emph{One has $\bm{y}^{(t)}_{>m} = \bm{0}$ for all $t=1,\ldots,T$}
\item[(b)] \emph{One has $\delta^2 \sum_{t=1}^T S(\bm{y}^{(t)}_{\leq m}) \leq \frac{C}{10}$}
\item[(c)] \emph{One has $\|\bm{Z}\|_2^2 \leq 5\alpha^2 m$}
\item[(d)] \emph{One has $\|\bm{Y}\|_2^2 \leq 5m$} 
\end{enumerate*}
{\bf Proof of claim.} By Corollary 5, and recalling that $m \ge \max\Big\{100,\frac{4}{\alpha^2}, \frac{1}{\eps^2}, \frac{2}{\alpha \varepsilon}\Big\}$, we can bound $D^2 = \frac{4\eps^2}{\alpha^2 m^2} \le \frac{1}{m}$ and $T = \frac{1}{\delta^2} \le 25m^9$, so that the failure probability for (a) is bounded by 
$T \cdot 2^{-m} \leq 25m^9 \cdot 2^{-m}   < \frac{1}{100}$. For (b), note that for every step $t$, the conditional expectation of $S(\bm{y}^{(t)}_{\leq m})$ is nonpositive, and $|S(\bm{y}^{(t)}_{\leq m})| \leq 4Dm^4$. 
Then using Azuma's inequality, one has $\Pr[\delta^2 \sum_{t=1}^T S(\bm{y}_{\leq m}^{(t)}) > \frac{C}{10}] \leq \exp( -\frac{1}{2} \cdot (\frac{C}{10} \delta)^2/(\delta^2 \cdot 4Dm^4)^2) \leq \frac{1}{20}$ since $C = \frac{10\eps}{\alpha} \ge 10\alpha \eps \ge \frac{20}{m}$. 

For $(c)$, note that $\E[\|\bm{Z}\|_2^2] = \sum_{t=1}^T \tr[\delta^2 \cdot (\bm{I}_m-\bm{X}^{(t)})] \leq \alpha^2 m$, so by Markov's inequality  $\Pr[\|\bm{Z}\|_2^2 > 5\alpha^2 m] \le \frac{1}{5}$. Similarly, $\E[\|\bm{Y}\|_2^2] \le m$, so $\|\bm{Y}\|_2^2 > 5m$ with probability at most $\frac{1}{5}$. The total failure probability is therefore at most $\frac{1}{100}+\frac{1}{20}+\frac{2}{5} < \frac{1}{2}$. \hfill \qedclaim

If the events in the claim hold, we have $\|\bm{Z}\|_2 \leq \alpha \sqrt{5m}, \bm{A}_{1.1C,D}(\bm{Y}) \succ 0$ and
\[\sum_{i=1}^m Y_i\bm{A}_i \preceq (1.1C+D\|\bm{Y}\|_2^2) \cdot \bm{I}_{2n} \preceq \Big(1.1 \cdot \frac{10\varepsilon}{\alpha} + \frac{10\varepsilon}{\alpha}\Big) = \frac{21\varepsilon}{\alpha} \cdot \bm{I}_{2n}.\] 

It remains to finish the arguments behind the proof strategy. By a slight abuse of notation, let $\gamma_m(\bm{x}) := \frac{1}{(2\pi)^{m/2}}e^{-\|\bm{x}\|_2^2/2}$ be the density of the Gaussian at a point $\bm{x}$. We define $p(\bm{x})$ as the conditional probability that the properties (a)-(d) are satisfied, conditioned on the event that $\bm{Y}+\bm{Z}=\bm{x}$. 
Then our reasoning above has proven that 
$\int_{\setR^m} \gamma_m(\bm{x}) \cdot p(\bm{x}) d\bm{x} \geq \frac{1}{2}$. Now define the set $Q := \{ \bm{x} \in \setR^m \mid p(\bm{x}) > 0\}$. As $0 \leq p(\bm{x}) \leq 1$ we must have
 $\gamma_m(Q) \geq \frac{1}{2}$. By construction, for every $\bm{x} \in Q$, there is at least one witness outcome $\bm{Y}+\bm{Z} = \bm{x}$
so that $\bm{Y} \in \frac{21 \varepsilon}{\alpha} K$ and $\|\bm{Z}\|_2 \leq \alpha \sqrt{5m}$. Then a slight reparametrization of $\alpha' := \sqrt{5} \alpha$ gives the claim 
as $21\sqrt{5} < 50$.

One final detail is that in Lemma 12 we assume $m \ge \frac{10}{\alpha^2}$. We now deal with smaller values of $\alpha$. Recall from the proof of Claim I that $\|\sum_{i=1}^m x_i \bm{A}_i \|_{\textrm{op}} \le \|\bm{x}\|_{\infty}$ for any $\bm{x} \in \setR^m$. In particular since $\|\bm{x}\|_{\infty} \le \|\bm{x}\|_2$ we have $\eps B^m_2 \subseteq K$, so $B^m_2 \subseteq \sqrt{n} B^m_2 \subseteq \sqrt{m} K$ and we find, say for $\alpha = \frac{7}{\sqrt{m}}$,
 $\gamma_m(15 \sqrt{m} K) \ge \gamma_m\Big((\frac{50}{\alpha} + \alpha m) K\Big) \ge \gamma_m \Big(\frac{50}{\alpha} K + \alpha \sqrt{m} B^m_2\Big) \ge \frac{1}{2}$. The conclusion is that Theorem 1 for $m < \frac{10}{\alpha^2}$ holds as we even have $\gamma_m(\frac{50}{\alpha} K) \ge \gamma_m (15 \sqrt{m} K) \ge \frac{1}{2}$.
\end{proof}

\section{Analysis of a single step}

In this section, we prove Lemma~\ref{lem:PotentialFunctionUpdateInOneIteration} and some variants that will be needed later.

\begin{proof}[Proof of Lemma~\ref{lem:PotentialFunctionUpdateInOneIteration}]
To simplify notation, we abbreviate matrices 
\[
\bm{A} := \bm{A}_{C,D}(\bm{x}), \quad  \tilde{\bm{B}} :=  \sum_{i=1}^m y_i\bm{A}_i, \quad \textrm{and} \quad \bm{B} := \tilde{\bm{B}} - \delta (D\|\bm{y}\|_2^2  + S(\bm{y})) \bm{I}_n.
\] 
Next, we define an index set
\[
\I := \Big \{i \in [m]: \tr\big[\bm{A}^{-1} |\bm{A}_i|\big] \leq \frac{2}{\alpha^2 m} \cdot \tr\big[\bm{A}^{-1}\big]\Big \}
\]
Here $[m] \setminus \I$ are the ``dangerous'' indices in the sense that updating $\bm{x}$ in these coordinates might disproportionally
change the potential function.
Note that by Markov's inequality, we have $|\I| \geq (1 - \frac{\alpha^2}{2})m$. Consider the subspace
\[
H :=\Big\{\bm{y} \in \setR^m : \left<\bm{x},\bm{y}\right> = 0, \sum_{i=1}^m y_i\cdot \tr[\bm{A}^{-2}\bm{A}_i]=0,\sum_{i=1}^m y_i\cdot \tr[\bm{A}^{-3}\bm{A}_i]=0, y_i = 0 \; \forall i \notin \I \Big\}
\] 
so that $\dim(H) \ge |\I| -3 \ge (1-\alpha^2)m$ for $m \ge \frac{10}{\alpha^2}$. Further, $\dim(H) \ge |\I| - 3 \ge 0.47m$ for $m \ge 100$. 
We choose $\bm{X}$ so that $N(\bm{0}, \bm{X})$ is the standard Gaussian restricted to $H$.

The remaining proof is organized in 4 claims, where Claim I-III justify that the 
Taylor approximation is well behaved while Claim IV contains a very crucial upper bound.
We begin by showing a rather crude upper bound on $|S(\bm{y})|$ for $\|\bm{y}\|_2 \le m$. \\
{\bf Claim I.} \emph{For every $\bm{y}$ with $\|\bm{y}\|_2 \leq m$, one has $|S(\bm{y})| \leq \frac{2m}{\delta}$, $\|\bm{B}\|_{\textrm{op}} \leq 4m$ and  $\|\delta \bm{A}^{-1} \bm{B}\|_{\textrm{op}} < \frac{1}{2}$.}  \\
{\bf Proof of Claim I.}
Note that in order for the potential functions
$\Phi_{C + \delta^2 S(\bm{y}),D}(\bm{x} + \delta \bm{y})$ and $\Phi_{C,D}(\bm{x})$ to be identical, we know that the difference matrix
\begin{align*} \bm{A}_{C+\delta^2 S(\bm{y}),D}(\bm{x}+\delta \bm{y})-\bm{A}_{C,D}(\bm{x}) &= \delta^2 (D\|\bm{y}\|_2^2 + S(\bm{y})) \cdot \bm{I}_n - \delta \sum_{i=1}^m y_i\bm{A}_i \\ 
&\preceq \delta^2 (Dm^2 + S(\bm{y})) \cdot \bm{I}_n + \delta \|\bm{y}\|_{\infty} \underbrace{\sum_{i=1}^m |\bm{A}_i|}_{\preceq \bm{I}_n} \\&\preceq \left(\delta D m^2 + \delta S(\bm{y}) + m\right) \cdot \delta \cdot \bm{I}_n   
\end{align*} 
must have one eigenvalue at least 0 and one eigenvalue at most $0$. 
There would be no positive eigenvalues if $\delta S(\bm{y}) < -2m < -\delta Dm^2 - m$, and similarly no negative eigenvalues if $\delta S(\bm{y}) > 2m$. Hence we conclude $|S(\bm{y})| \leq \frac{2m}{\delta}$.
This bound is good enough to show that
\[
\|\bm{B}\|_{\textrm{op}} \le \underbrace{\Big\|\sum_{i = 1}^m y_i \bm{A}_i\Big\|_{\textrm{op}}}_{\le m} + \delta \Big(D\|\bm{y}\|^2_2 + \frac{2m}{\delta}\Big) \le 3m + \delta Dm^2 < 4m.
\]
Since $\|\bm{A}^{-1}\|_{\textrm{op}} \le \Phi_{C,D}(\bm{x}) \leq \frac{Dm^2}{10}$, it follows $\|\delta \bm{A}^{-1} \bm{B}\|_{\textrm{op}} \le \delta \|\bm{A}^{-1}\|_{\textrm{op}} \| \bm{B}\|_{\textrm{op}} < \frac{1}{2}$. \hfill \qedclaim

Now we can apply the matrix Taylor approximation from Lemma~\ref{lem:MatrixTaylorApprox} and use that 
for every $\bm{y} \in H$ with $\|\bm{y}\|_2 \leq m$, there exists some $|c| \le 2$ such that the difference in the potential function is
\begin{eqnarray*}
 0 &\stackrel{\textrm{Def }S(\bm{y})}{=}& \Phi_{C+\delta^2 S(\bm{y}),D}(\bm{x} + \delta \bm{y}) - \Phi_{C,D}(\bm{x})  \\
&=& \tr\left[(\bm{A}-\delta \bm{B})^{-1}\right] - \tr\left[\bm{A}^{-1}\right] \\
 &\stackrel{\textrm{Lem~\ref{lem:MatrixTaylorApprox}}}{=}& \delta \cdot \tr[\bm{A}^{-1}\bm{B}\bm{A}^{-1}] + c\delta^2 \cdot \tr[\bm{A}^{-1}\bm{B}\bm{A}^{-1}\bm{B}\bm{A}^{-1}]  \\
&=& -\delta^2 (D\|\bm{y}\|_2^2+S(\bm{y}))\cdot \tr[\bm{A}^{-2}\bm{I}_n] + \delta \underbrace{\sum_{i=1}^m y_i \tr[\bm{A}^{-1}\bm{A}_i\bm{A}^{-1}]}_{=0} + c\delta^2 \cdot \tr[\bm{A}^{-1}\bm{B}\bm{A}^{-1}\bm{B}\bm{A}^{-1}] \\
 &\stackrel{\bm{y} \in H}{=}& \delta^2 \Big( -(D\|\bm{y}\|_2^2+S(\bm{y})) \cdot \tr[\bm{A}^{-2}] + c \cdot \tr[\bm{A}^{-1}\bm{B}\bm{A}^{-1}\bm{B}\bm{A}^{-1}] \Big). \hspace{3cm} (**)
\end{eqnarray*}
Observe that in the last equation we have conveniently used that due to the linear constraints defining $H$,
we have  $\tr[\bm{A}^{-1}\tilde{\bm{B}}\bm{A}^{-1}]=0$ for all $\bm{y} \in H$.
Now we can show that the quantity $S(\bm{y})$ is a lot smaller than we have proven so far --- in fact its maximum
length is independent of the step size $\delta$: \\
{\bf Claim II.} \emph{For every $\bm{y} \in H$ with $\|\bm{y}\|_2 \leq m$ one has $|S(\bm{y})| \leq 4Dm^4$.} \\
{\bf Proof of Claim II.} We rearrange $(**)$ for $S(\bm{y})$ and obtain
\[
 |S(\bm{y})| \leq \frac{|c| \cdot \tr[\bm{A}^{-1}\bm{B}\bm{A}^{-1}\bm{B}\bm{A}^{-1}]}{\tr[\bm{A}^{-2}]} + D\underbrace{\|\bm{y}\|_2^2}_{\leq m^2}   
\leq 2\cdot \|\bm{A}^{-1}\|_{\textrm{op}} \cdot \|\bm{B}\|^2_{\textrm{op}} + Dm^2 \le 4Dm^4,
\]
using the estimates $\|\bm{B}\|_{\textrm{op}} \le 4m$ and $\|\bm{A}^{-1}\|_{\textrm{op}} \le \frac{Dm^2}{10}$.  \hfill \qedclaim

Next, we justify that $\tr[\bm{A}^{-1}\bm{B}\bm{A}^{-1}\bm{B}\bm{A}^{-1}] \approx \tr[\bm{A}^{-1}\tilde{\bm{B}}\bm{A}^{-1}\tilde{\bm{B}}\bm{A}^{-1}]$ up to lower order terms. 


\noindent {\bf Claim III.} \emph{For any $\bm{y} \in H$ with $\|\bm{y}\|_2 \leq m$ one has} 
\[
 \left|\tr[\bm{A}^{-1}\bm{B}\bm{A}^{-1}\bm{B}\bm{A}^{-1}]- \tr[\bm{A}^{-1}\tilde{\bm{B}}\bm{A}^{-1}\tilde{\bm{B}}\bm{A}^{-1}]\right| \leq \delta^2 \cdot \tr[\bm{A}^{-2}] \cdot \frac{5}{2}D^3m^{10}   
\]
{\bf Proof of Claim III.} Since $\bm{B} = \tilde{\bm{B}} - \delta (D\|\bm{y}\|_2^2  + S(\bm{y})) \bm{I}_n$, the difference in the left side equals

\[ 
\Big|-2\delta (D \|\bm{y}\|_2^2 + S(\bm{y}))\underbrace{\tr[\bm{A}^{-3} \tilde{\bm{B}}]}_{= 0}+ \delta^2 \tr[\bm{A}^{-3}](D \|\bm{y}\|_2^2 + S(\bm{y}))^2\Big| \le \delta^2 \tr[\bm{A}^{-2}] \cdot \frac{Dm^2}{10} \cdot (5Dm^4)^2
\] 
Here we use $\tr[\bm{A}^{-3}] \le \frac{Dm^2}{10} \cdot \tr[\bm{A}^{-2}]$, as well as $(D \|\bm{y}\|_2^2 + S(\bm{y}))^2 \le (5Dm^4)^2$. 
In particular we have also made use of the linear constraint $\sum_{i=1}^m y_i\tr[\bm{A}^{-3}\bm{A}_i]=0$ in the choice of the subspace $H$. \qedclaim

Now we prove the central core of this theorem: in expectation for a Gaussian $\bm{y}$ from the subspace $H$, 
the quadratic term $\tr[\bm{A}^{-1}\tilde{\bm{B}}\bm{A}^{-1}\tilde{\bm{B}}\bm{A}^{-1}]$ is bounded by a term that we can offset in the potential function by the
length increase of $\bm{x}$. \\
{\bf Claim IV.} \emph{One has $\E_{\bm{y} \sim N_{\leq m}(\bm{0},\bm{X})}\big[\tr[\bm{A}^{-1}\tilde{\bm{B}}\bm{A}^{-1}\tilde{\bm{B}}\bm{A}^{-1}]\big] \leq \frac{2}{\alpha^2 m} \tr[\bm{A}^{-1}] \cdot \tr[\bm{A}^{-2}]$.}  \\
{\bf Proof of Claim IV.} The argument for this claim needs some care, as we have in general $\E[y_iy_j] \neq 0$
since we draw $\bm{y}$ from a subspace $H$. 
We abbreviate $\bm{W}_i := \bm{A}^{-1/2}\bm{A}_i\bm{A}^{-1} \in \setR^{n \times n}$ (note that these matrices will in general not be symmetric). Then
\begin{eqnarray*}
 \E_{\bm{y} \sim N_{\leq m}(\bm{0},\bm{X})}\Big[\tr[\bm{A}^{-1}\tilde{\bm{B}}\bm{A}^{-1}\tilde{\bm{B}}\bm{A}^{-1}]\Big] &\stackrel{(i)}{=}& \E_{\bm{y} \sim N_{\leq m}(\bm{0},\bm{X})}\Big[\sum_{i \in \I} \sum_{j \in \I} y_iy_j\tr\left[\bm{A}^{-1}\bm{A}_i\bm{A}^{-1}\bm{A}_j\bm{A}^{-1}\right]\Big] \\
&=& \E_{\bm{y} \sim N_{\leq m}(\bm{0},\bm{X})}\Big[\sum_{i \in \I} \sum_{j \in \I} y_iy_j \left<\bm{W}_i,\bm{W}_j\right>_F\Big] \\
&=& \E_{\bm{y} \sim N_{\leq m}(\bm{0},\bm{X})}\Big[\Big\| \sum_{i \in \I} y_i \bm{W}_i\Big\|_F^2\Big] \\
&\stackrel{(ii)}{\leq}& \sum_{i \in \I} \|\bm{W}_i\|_F^2 = \sum_{i \in \I}\tr[\bm{A}^{-1}\bm{A}_i\bm{A}^{-1}\bm{A}_i\bm{A}^{-1}] \\
&\stackrel{(iii)}{=}& \sum_{i \in \I}\tr[\bm{A}^{-2}\bm{A}_i\bm{A}^{-1}\bm{A}_i] 
\stackrel{\textrm{Lem~\ref{lem:TraceOfProductOfMatrices}}}{\leq} \sum_{i \in \I} \tr\big[\bm{A}^{-2}|\bm{A}_i|\big] \cdot \tr\big[\bm{A}^{-1}|\bm{A}_i|\big]   \\
&\stackrel{(iv)}{\leq}& \frac{2}{\alpha^2 m} \tr[\bm{A}^{-1}] \sum_{i \in \I} \tr\big[\bm{A}^{-2}|\bm{A}_i|\big]  \\
&=& \frac{2}{\alpha^2 m} \tr[\bm{A}^{-1}] \tr\Big[\bm{A}^{-2} \cdot \underbrace{\sum_{i \in \I} |\bm{A}_i|}_{\preceq \bm{I}_n} \Big] \\
&\stackrel{\bm{A}^{-2} \succ 0}{\leq}& \frac{2}{\alpha^2 m} \tr[\bm{A}^{-1}] \cdot \tr[\bm{A}^{-2}]. 
\end{eqnarray*}
In $(i)$, we use that $y_i=0$ for $i \notin \I$. In $(ii)$ we use Lemma~\ref{lem:PropertiesGaussianFromSubspace} with the subtlety that replacing $\bm{y} \sim N(\bm{0},\bm{X})$ by the capped sample $\bm{y} \sim N_{\leq m}(\bm{0},\bm{X})$ can only decrease the length $\|\sum_{i \in \I} y_i\bm{W}_i\|_F^2$. In $(iii)$ we use cyclicity of the trace and in $(iv)$ we use that we have selected the indices $\I$ so that $\tr[\bm{A}^{-1}|\bm{A}_i|] \leq \frac{2}{\alpha^2 m} \tr[\bm{A}^{-1}]$ for any $i \in \I$.  \hfill \qedclaim

Now we have everything to finish the analysis. Taking expectation over $\bm{y} \sim N_{\leq m}(\bm{0},\bm{X})$ on both sides of $(**)$ gives
\begin{eqnarray*}
0 &\stackrel{(**)}{=}&  -(D\E[\|\bm{y}\|_2^2]+\E[S(\bm{y})]) \cdot \tr[\bm{A}^{-2}] + c \cdot \E[\tr[\bm{A}^{-1}\bm{B}\bm{A}^{-1}\bm{B}\bm{A}^{-1}]] \\
&\stackrel{\textrm{Claim III}}{\leq}& -(D\E[\|\bm{y}\|_2^2] + \E[S(\bm{y})]) \cdot \tr[\bm{A}^{-2}] + 2\E[\tr[\bm{A}^{-1}\tilde{\bm{B}}\bm{A}^{-1}\tilde{\bm{B}}\bm{A}^{-1}]] + \tr[\bm{A}^{-2}] \cdot \frac{D}{5} \\
&\stackrel{\textrm{Claim IV}}{\leq}& \Big(-0.45Dm  -\E[S(\bm{y})] + \frac{4}{\alpha^2 m} \tr[\bm{A}^{-1}] + \frac{D}{5} \Big) \cdot \tr[\bm{A}^{-2}] \\
&\stackrel{m \ge 20}{\leq}& \Big(-0.44Dm  -\E[S(\bm{y})] + \frac{4}{\alpha^2 m} \tr[\bm{A}^{-1}]  \Big) \cdot \tr[\bm{A}^{-2}] 
\end{eqnarray*}
In the first inequality we have used Claim III with the fact that $\delta^2 \leq \frac{1}{25D^2 m^{10}}$.
Here we also use that by Corollary 5 one has
$ 
\E[\|\bm{y}\|_2^2] \geq \dim(H) \cdot \big(1-2^{-\dim(H)}\big) \geq 0.45m
$ as $\dim(H) \ge 0.47 m$ and $m \geq 10$. 
Combining the two above inequalities, we conclude
\[
 \E[S(\bm{y})] \leq \frac{4}{\alpha^2 m} \tr[\bm{A}^{-1}] - 0.44D m \leq \frac{4}{\alpha^2 m} \tr[\bm{A}^{-1}] - \frac{4D m}{10} \le 0,
\]
making use of the assumed bound on $\Phi_{C,D}(\bm{x})$. 

It remains to argue $\tilde{\bm{A}} := \bm{A}_{C + \delta^2 S(\bm{y}), D} \succ 0$. Recall from the proof of Claim I we have 
\[
\tilde{\bm{A}} - \bm{A} = \delta^2 (D\|\bm{y}\|_2^2 + S(\bm{y})) \cdot \bm{I}_n - \delta\sum_{i=1}^m y_i \bm{A}_i \succeq (\delta Dm^2 + \delta S(\bm{y}) - m)\cdot \delta \cdot \bm{I}_n \succeq -3m\cdot \delta \cdot \bm{I}_n,
\]
where we have used $\delta S(\bm{y}) \ge -2m$. Remark that the least eigenvalue of $\bm{A}$ is at least $\frac{10}{Dm^2 \alpha^2}$. It follows that the least eigenvalue of $\tilde{\bm{A}}$ is at least $\frac{10}{Dm^2} - \frac{3}{5Dm^4} = \frac{10m^2 - 0.6}{Dm^4} > 0.$

\end{proof}

For later, it will be useful to consider the intersection of $K$ with linear constraints that force a constant
fraction of variables to be $0$:

 \begin{lemma} \label{lem:MainResultGaussianExpansionOfKJ}
Let $\bm{A}_1,\ldots,\bm{A}_m \in \setR^{n \times n}$ be positive semidefinite matrices with $\sum_{i=1}^{m} |\bm{A}_i| \preceq \bm{I}_n$. Let $\alpha, \beta, \eps \in (0, 1)$ with $m = \frac{n}{\eps^2}$, $\frac{1}{5} \ge \alpha^2 \ge 5\beta$ and $J \subset [m]$ with $|J| \le \beta m$. Then the set 
$$K := \Big\{ \bm{x} \in \setR^m \mid \Big\| \sum_{i=1}^m x_i \bm{A}_i \Big\|_{\textrm{op}} \leq \eps \Big\}$$ 
also satisfies $\gamma_m(\frac{50}{\alpha} K(J) + \alpha \sqrt{m} B_2^m) \geq \frac{1}{2}$, where $K(J) = K \cap \{\bm{x} : x_j = 0 \ \forall j \in J\}.$
\end{lemma}
\begin{proof}
We can reuse the proof of Theorem~\ref{thm:MainResultGaussianExpansionOfK} unchanged, but we 
revisit the proof of Lemma~\ref{lem:PotentialFunctionUpdateInOneIteration} and in particular the choice of the
subspace $H$. Suppose we modify the definition of $H$ and add the linear constraints $y_j = 0$ for all $j \in J$. 
The dimension of the subspace will still be  $\dim(H) \geq \big(1 - \frac{\alpha^2}{2} - \beta\big) m - 3 \ge (1-\alpha^2) m$ for $m \ge \frac{10}{\alpha^2}$ as $\beta \le \frac{\alpha^2}{5}$. The dimension is also at least $(0.47 - \beta)m $ for $m \ge 100$.
The remaining proof of Lemma~\ref{lem:PotentialFunctionUpdateInOneIteration}  applies as we still have $\E[S(\bm{y})] \le
\frac{4}{\alpha^2 m } \tr[\bm{A}^{-1}] - (0.44 - \beta) Dm \le (0.04m - \beta)Dm \le 0.
$
\end{proof}
The attentive reader may have noticed that the proof of Theorem~\ref{thm:MainResultGaussianExpansionOfK}
allows to handle a concentration that should be a lot tighter than just the factor of 1/2 that we obtained. 
But it is a well-known insight that Gaussian measures can be boosted using the Gaussian Isoperimetric inequality.
\begin{lemma}
With the notation from Lemma~\ref{lem:MainResultGaussianExpansionOfKJ}, for any $\delta > 0$ we have 
\[
\gamma_m\Big(\frac{50}{\alpha} K(J) + (\alpha + \delta) \sqrt{m} B_2^m\Big) \ge 1 - e^{-\delta^2m/2}.
\]
\end{lemma}

\begin{proof}
It suffices to apply the \emph{Gaussian Isoperimetric inequality} with Lemma 13 to get 
\[
\gamma_m\Big(\frac{50}{\alpha} K(J) + (\alpha + \delta) \sqrt{m} B_2^m\Big) \ge 1 - \int_{\delta \sqrt{m}}^{\infty} \frac{1}{\sqrt{2\pi}} e^{-x^2/2} dx \ge 1 - e^{-\delta^2m/2}.
\]
\end{proof}

\section{Mean width and Gaussian measure}

One of the standard quantities that are studied in the context of convex bodies $K$ is the 
mean width $w(K) = \E_{\bm{a} \in S^{n-1}}[\max_{\bm{x},\bm{y} \in K} |\left<\bm{a},\bm{x}-\bm{y}\right>|]$. The wonderful textbook of Artstein-Avidan, Giannopoulos and Milman~\cite{AsymptoticGeometricAnalysisBook2005} contains many applications.
Additionally, the analysis of Eldan and Singh~\cite{DBLP:journals/rsa/EldanS18} of a modification of the algorithm of \cite{ConstructiveDiscrepancy-Rothvoss-FOCS2014} also makes use
of the width of a body.
We can prove that the mean width of the body $K$ arising in our spectral setting is indeed high. 
\begin{theorem}
Let $\bm{A}_1,\ldots,\bm{A}_m \in \setR^{n \times n}$ be symmetric matrices with $\sum_{i=1}^{m} |\bm{A}_i| \preceq \bm{I}_n$ and select $\varepsilon \in (0, 1)$ so that $m = \frac{n}{\eps^2} \ge 100$. Then the set
\[
 K := \Big\{ \bm{x} \in \setR^m \mid \Big\| \sum_{i=1}^m x_i \bm{A}_i \Big\|_{\textrm{op}} \leq \eps \Big\}
\]
has mean width $w(K) \geq \Omega(\sqrt{m})$.
\end{theorem}
\begin{proof}
Let $\alpha>0$ be a small constant that we determine later. Consider the body $Q := \frac{50}{\alpha}K + \alpha \sqrt{m}B_2^m$. Then by Theorem~\ref{thm:MainResultGaussianExpansionOfK} we know that $\gamma_m(Q) \geq \frac{1}{2}$. 
We want to first show that $Q$ has a high mean width. 
One can check that $\Pr_{\bm{y} \sim N(\bm{0},\bm{I}_m)}[\|\bm{y}\|_2 < 0.9\sqrt{m}] \leq \frac{1}{4}$ for $m \geq 100$. Then, 
\begin{eqnarray*}
 w(Q) &\geq& \E_{\bm{y} \in N(\bm{0},\bm{I}_m)}\Big[\max\Big\{ \Big<\frac{\bm{y}}{\|\bm{y}\|_2},\bm{x}\Big> : \bm{x} \in Q\Big\}\Big] \geq \E_{\bm{y} \in N(\bm{0},\bm{I}_m)}\Big[ \Big<\frac{\bm{y}}{\|\bm{y}\|_2},\bm{y}\Big> \cdot \bm{1}_{\bm{y} \in Q}\Big] \\
&\geq& 0.9\sqrt{m} \cdot \Pr_{\bm{y} \in N(\bm{0},\bm{I}_m)}\big[\|\bm{y}\|_2 > 0.9\sqrt{m}\textrm{ and }\bm{y} \in Q\big] \geq 0.9\sqrt{m} \cdot \Big(1 - \frac{1}{4}-\frac{1}{2}\Big) > \frac{1}{5}\sqrt{m}.
\end{eqnarray*}
Observe that the mean width is additive and scales with the body, hence
\[
\frac{1}{5} \sqrt{m} < w(Q) = w\Big(\frac{50}{\alpha}K + \alpha \sqrt{m}B_2^m\Big) = \frac{50}{\alpha} \cdot w(K) + \alpha \sqrt{m} \cdot \underbrace{w(B_2^m)}_{=2}
\]
This can be rearranged to 
\[
 w(K) > \frac{\alpha}{50} \cdot \Big(\frac{1}{5}\sqrt{m}-2\alpha \sqrt{m}\Big) \stackrel{\alpha := \frac{1}{20}}{=} \frac{1}{10000}\sqrt{m}
\]
\end{proof} 
Note that one could certainly obtain a tighter constant using heavier machinery. In particular Urysohn's inequality states
that the mean width of any body is at least that of an Euclidean ball with equal volume.

We also conjecture that the following bound on the Gaussian measure holds: 
\begin{conjecture} \label{conj:GaussianMeasureForKatLeast2m}
Using the same notation from Theorem 1, we have $\gamma_m(K) \ge 2^{-cm}$ for a universal constant $c>0$.
\end{conjecture}
Note that Conjecture~\ref{conj:GaussianMeasureForKatLeast2m} would also imply Theorem~\ref{thm:MainResultGaussianExpansionOfK}. In fact, 
from the Gaussian Isoperimetric Inequality one can derive that any set $K$ with $\gamma_m(K) \geq 2^{-cm}$ also satisfies $\gamma_m(K + 4\sqrt{cm} \cdot B_2^m) \geq \frac{1}{2}$. We comment that a lower bound of  $\gamma_m(K) \geq 2^{-cm}$ would be best possible in general. To see this, consider the case where $\bm{A}_i = \bm{e}_i \bm{e}_i^\top$ for $i \le n$ and $\bm{0}$ otherwise, so that $K = \{\bm{x} \in \setR^m: |x_i| \le \eps, i \le n\}$ 
which has Gaussian measure $\Big(\frac{2m}{n
}\Big)^{-cn}$, indeed $2^{-cm}$ for $m = n$.
The best lower bound on $\gamma_m(K)$  that we are aware of comes from $[-\eps, \eps]^m \subseteq K$, so that we get $\gamma_m(K) \ge \gamma_m([-\eps, \eps]^m)= \Big(\frac{2m}{n
}\Big)^{-cm}$. One difficulty for proving a lower bound on $\gamma_m(K)$ is that the Gaussian measure is in some sense a more brittle property than mean width --- the intersection of $K$ with a single hyperplane brings the measure down to 0 while the mean width is little affected. Of course, the body $K$ in our setting is full-dimensional but it is less clear that it is sufficiently fat in enough directions. Another observation is that we have indeed proven that  $\gamma \Big(\frac{50}{\alpha} K + \alpha \sqrt{m} B^m_2\Big) \ge \frac{1}{2}$ for the whole \emph{range} of $\alpha>0$.
In fact, we do not know any convex symmetric body $K \subseteq \setR^m$ with measure $\gamma_m(K) = \log(m)^{-cm}$, such that the conclusion of Theorem 1 holds, that is, $\gamma_m \Big(\frac{50}{\alpha} K + \alpha \sqrt{m} B^m_2\Big) \ge \frac{1}{2}$ for all $\alpha > 0$. 
As an exercise, it is not hard to verify any \emph{cylinder} of the form $C = \{x_1^2 + \dots + x_d^2 \le r\}$ for $d \le m$ and $r > 0$ that satisfies the conclusion of Theorem~\ref{thm:MainResultGaussianExpansionOfK} will indeed have $\gamma_m (C) \ge 2^{-cm}$.

\section{From high mean width to efficient algorithms}

In this section, we prove the correctness of the spectral sparsification algorithm from Section~\ref{sec:Contribution}.  The algorithm runs logarithmically many iterations of a routine due to \cite{ConstructiveDiscrepancy-Rothvoss-FOCS2014}. Consider an arbitrary symmetric convex set $K \subseteq \setR^m$ with measure $\gamma_m(K) \geq 2^{-cm}$ for a small enough constant $c>0$. Then one can sample a random Gaussian $\bm{x}^* \sim N(\bm{0},\bm{I}_m)$ and compute the point $\bm{y}^* \in K \cap [-1,1]^m$ that minimizes the distance $\|\bm{x}^*-\bm{y}^*\|_2$. Then \cite{ConstructiveDiscrepancy-Rothvoss-FOCS2014} shows that 
with probability $1-2^{-\Omega(m)}$, the point $\bm{y}^*$ has at least  $\beta m$ many entries in $\{ -1,1\}$, where $\beta$ is a small enough constant. 
We reproduce a picture of~\cite{ConstructiveDiscrepancy-Rothvoss-FOCS2014}:
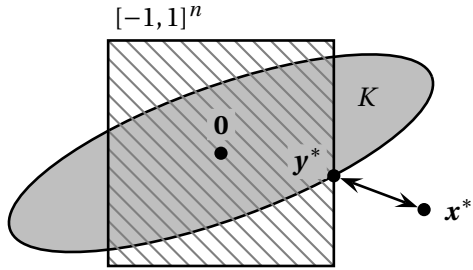
\begin{center} 
\psset{unit=1.5cm}
\begin{pspicture}(-2.4,-1)(2.4,1.2)
\rput[c]{20}(0,0){\psellipticarc[fillstyle=solid,fillcolor=lightgray](0,0)(2,0.6){0}{-0.1}}
\pspolygon[fillstyle=vlines,fillcolor=lightgray,hatchcolor=gray](-1,-1)(1,-1)(1,1)(-1,1)
\rput[c](1.3,0.5){\psframebox[framesep=2pt,fillstyle=solid,fillcolor=lightgray,linestyle=none]{$K$}}
\cnode*(0,0){2.5pt}{origin} \nput[labelsep=2pt]{90}{origin}{\psframebox[fillstyle=solid,fillcolor=lightgray,framesep=2pt,linestyle=none]{$\bm{0}$}}
\cnode*(1.8,-0.5){2.5pt}{x} \nput{0}{x}{$\bm{x}^*$}
\cnode*(1,-0.2){2.5pt}{y} \nput[labelsep=0pt]{150}{y}{\psframebox[fillstyle=solid,fillcolor=lightgray,framesep=1pt,linestyle=none]{$\bm{y}^*$}}
\rput[l](-0.95,1.2){$[-1,1]^n$}
\ncline[arrowsize=6pt,linewidth=1pt]{<->}{x}{y}
\end{pspicture}
\end{center}
The paper uses the property
 $\gamma_m(K) \geq 2^{-cm}$ to derive that in particular for every index set $J$ with $|J| \leq \beta m$, the random point $\bm{x}^*$ would be far from $K$ intersected with coordinate slabs $|x_i| \leq 1$ for all $i \in J$. While in our setting, we do not know whether the premise of $\gamma_m(K) \geq 2^{-cm}$ is true, we prove that the intermediate property is satisfied (even for intersections with hyperplanes $x_i=0$ instead of slabs). 
We provide the details of the analysis below:

\begin{lemma}
Let $\alpha, \beta \in (0, 1)$ with $\alpha \le \frac{1}{5}$ and $\theta > \theta' := \frac{3}{2} \beta \log (\frac{1}{\beta})$. Suppose $K \subseteq \setR^m$ is a symmetric convex body with $\gamma_m(K(J) + \alpha \sqrt{m} B^m_2) \ge 1 - e^{-\theta m}$ for all $J \subseteq [m]$ with $|J| \le \beta m$. Sample $\bm{x}^* \sim N(\bm{0}, \bm{I}_m)$ and let $\bm{y}^*$ be the point in $K \cap [-1,1]^m$ that minimizes $\|\bm{x}^* - \bm{y}^*\|_2$. Then with probability $1 - 2^{-\Omega(m)}$, $\bm{y}^*$ has at least $\beta m$ coordinates $i$ with $y^*_i \in \{-1,1\}$. 
\end{lemma}

\begin{proof}
First note that $\Pr_{\bm{x}^* \sim N(\bm{0}, \bm{I}_m)}[|x^*_i| \ge 2] = 2 \int_2^\infty \frac{1}{\sqrt{2\pi}} e^{-t^2/2} dt > \frac{1}{25}$, so with probability $1 - 2^{-\Omega(m)}$ we have $d(\bm{x}^*, [-1,1]^m) > \sqrt{\frac{m}{25} \cdot (2-1)^2} = \frac{1}{5} \sqrt{m}$. Recall that $K(J) = K \cap \{ \bm{x} \in \setR^m \mid x_i=0 \; \forall i \in J\}$. We also define the set $K_{\textsc{Strips}}(J) := K \cap \{ \bm{x} \in \setR^m \mid |x_i| \leq 1 \; \forall i \in J\}$.
Consider the index set $J^* := \{i \in [m] : y^*_i \in \{-1,1\}\}$ and note that $d(\bm{x}^*, K \cap [-1,1]^m) = d(\bm{x}^*, K_{\textsc{Strips}}(J^*))$ since $J^*$ defines the tight constraints for $\bm{y}^*$.

 Since there are at most $e^{\theta' m}$ sets $J \subseteq[m]$ with $|J| \le \beta m$, using the union bound gives
 \[
\gamma_m\Big(\bigcup_{|J| \le \beta m} \Big(\setR^m \setminus (K(J) + \alpha \sqrt{m} B^m_2)\Big) \Big) \le \sum_{|J| \le \beta m} \gamma_m(\setR^m \setminus (K(J) + \alpha \sqrt{m} B^m_2)) \le e^{(\theta'- \theta) m}. 
\]
So with probability $1 - 2^{-\Omega(m)}$, one has
\[
d(\bm{x}^*,K(J^*)) \geq d(\bm{x}^*, K_{\textsc{Strips}}(J^*)) = d(\bm{x}^*, K \cap [-1,1]^m) \ge d(\bm{x}^*, [-1,1]^m) \ge \frac{1}{5} \sqrt{m}
\] 
whereas $d(\bm{x}^*, K(J)) \le \alpha \sqrt{m} \le \frac{1}{5} \sqrt{m}$ for all $J$ with $|J| \le \beta m$. It follows $|J^*| > \beta m$.
\end{proof}
More specifically for our spectral setting we can find fractional partial colorings with the following guarantee: 
\begin{corollary} \label{cor:FractPartialColoringForSpectralK}
Let $\bm{A}_1,\ldots,\bm{A}_m \in \setR^{n \times n}$ be symmetric matrices with $\sum_{i=1}^{m} |\bm{A}_i| \preceq \bm{I}_n$. Select $\eps \in (0,1)$ so that $m = \frac{n}{\eps^2} \ge 100$, and define the set  $$K := \Big\{ \bm{x} \in \setR^m \mid \Big\| \sum_{i=1}^m x_i \bm{A}_i \Big\|_{\textrm{op}} \leq \eps \Big\}.$$
There is a polynomial time algorithm that returns $\bm{y}^* \in 500K \cap [-1,1]^m$ with at least $\frac{m}{9000}$ coordinates $y^*_i$ in $\{-1,1\}$, with probability $1- 2^{-\Omega(m)}$.
\end{corollary}
\begin{proof}
It suffices to choose $\beta = \frac{1}{9000}$, $\alpha = \frac{15}{100}$ and $\delta = \frac{5}{100}$, so that $\frac{1}{5} \ge \alpha^2 \ge 5\beta$ and we may apply Lemma 14 to get a lower bound on the Gaussian measure. We also have $\frac{50}{\alpha} < 500$, $\alpha + \delta = \frac{1}{5}$ and $\frac{\delta^2}{2} > \frac{3}{2} \beta \log(\frac{1}{\beta})$, thus applying the previous lemma gives the corollary. Finally note that finding the point $\bm{y}^* \in 500K \cap [-1,1]^m$ that minimizes $\|\bm{x}^*-\bm{y}^*\|_2$
is a convex optimization problem and can be solved in polynomial time for example with the help
of the Ellipsoid method~\cite{DBLP:books/sp/GLS1988}. Here, for a given $\bm{y}^* \notin 500K \cap [-1,1]^m$, a separating hyperplane can be derived 
from the eigendecomposition of the matrix $\sum_{i=1}^m y_i^*\bm{A}_i$.
\end{proof}

Finally, we can prove Theorem 3 and give an analysis of the full algorithm from Section~\ref{sec:Contribution}.
The basic intuition is that we start with a weight vector $\bm{s} := (1,\ldots,1)$ so that $\sum_{i=1}^m s_i\bm{A}_i=\bm{I}_n$. Then 
in each iteration we find a partial coloring according to Corollary~\ref{cor:FractPartialColoringForSpectralK} and 
we use the partial coloring to update the weights so that at least a constant fraction of the weights drop to 0. 

\begin{proof}[Proof of Theorem 3]
Consider one iteration of the algorithm where the current weights are $\bm{s} \in \setR_{\geq 0}^m$. 
The body defined in step (3) is $K := \{ \bm{x} \in \setR^{\supp(\bm{s})} \mid \|\sum_{i \in \textrm{supp}(\bm{s})} x_is_i\bm{A}_i\|_{\textrm{op}} \leq 1000\tilde{\varepsilon} \}$.  
Hence, by Corollary~\ref{cor:FractPartialColoringForSpectralK} applied to matrices $\bm{A}_i' := \frac{s_i}{2} \bm{A}_i$, 
we know that after line (6), we  have a point $\bm{x}^* \in [-1,1]^{\textrm{supp}(\bm{s})}$ with $\|\sum_{i=1}^m x^*_i s_i\bm{A}_i\|_{\textrm{op}} \le 1000 \sqrt{\frac{n}{|\textrm{supp}(\bm{s})|}}$ and at least $\frac{1}{2} \cdot \frac{|\textrm{supp}(\bm{s})|}{9000}$ coordinates equal to $-1$. Thus, at line (7), $|\textrm{supp}(\bm{s})|$ is reduced by a factor of $\kappa:= 1-1/18000 < 1$. It follows that the algorithm terminates after $O(\log(\frac{\eps^2m}{n})) = O(\log m)$ loop iterations. Further, at each iteration, we add $\sum_{i=1}^m x^*_is_i\bm{A}_i$ to the matrix $\sum_{i=1}^m s_i\bm{A}_i$, which is originally $\bm{I}_n$. So by triangle inequality, at the end of the algorithm we have an additive error of at most 
\[
1000\sum_{t \ge 0} \sqrt{\frac{n}{\kappa^t m}} = O\Big(\sqrt{\frac{n}{m}}\Big) = O(\eps),   
\]
that is, $(1-O(\eps)) \bm{I}_n \preceq \sum_{i=1}^m s_i\bm{A}_i \preceq (1+O(\eps)) \bm{I}_n$. Note that the argument also provides that in every single iteration we had $\sum_{i=1}^m s_i\bm{A}_i \preceq 2\bm{I}_n$ for small enough $\eps > 0$, which justifies the application of Corollary~\ref{cor:FractPartialColoringForSpectralK} in the first place. The error probability is dominated by $2^{-\Theta(m_0)}$, where $m_0 \geq n$ is the support in the last iteration.
\end{proof}

\section{Open questions and conjectures}

We close this paper by presenting a range of open questions that did arise in the context of this work. We begin by reiterating a question that we had mentioned earlier in the form of a conjecture, but specialize it here to rank-1 to keep it as simple as possible: 
\begin{question}
Is it true that there is a universal constant $c>0$ so that for any vectors
 $\bm{v}_1,\ldots,\bm{v}_m \in \setR^n$ with $\sum_{i=1}^m \bm{v}_i\bm{v}_i^T = \bm{I}_n$, the body $K := \{ \bm{x} \in \setR^m \mid \|\sum_{i=1}^mx_i\bm{v}_i\bm{v}_i^\top\|_{\textrm{op}} \leq \varepsilon\}$ has measure $\gamma_m(K) \geq 2^{-cm}$, where $\varepsilon$ is chosen so that $m = \frac{n}{\varepsilon^2}$.
\end{question}
We also restate the conjecture popularized by Meka: 
\begin{question} [Matrix Spencer Conjecture]
Is it true that there is a universal constant $C > 0$ so that for all symmetric matrices $\bm{A}_1,\ldots,\bm{A}_n \in \setR^{n \times n}$ with $\|\bm{A}_i\|_{\textrm{op}} \leq 1$ for $i \in [n]$, there are signs $\bm{x} \in \{ -1,1\}^n$ satisfying $\|\sum_{i=1}^n x_i\bm{A}_i\|_{\textrm{op}} \leq C\sqrt{n}$.
\end{question}
A statement that would allow at least a good partial coloring can be
formalized as follows:  
\begin{question} \label{question:KForSumOfBoundedOpnorm-has-LargeMeasure}
Is it true that there is a universal constant $c > 0$ so that for all symmetric matrices $\bm{A}_1,\ldots,\bm{A}_n \in \setR^{n \times n}$ with $\|\bm{A}_i\|_{\textrm{op}} \leq 1$ for $i \in [n]$, $K := \{ \bm{x} \in \setR^n \mid \|\sum_{i=1}^n x_i\bm{A}_i\|_{\textrm{op}} \leq \sqrt{n} \}$ has Gaussian measure $\gamma_n(K) \geq 2^{-cn}$.
\end{question}
One can again ask an even weaker question that according to our experience might have a simpler answer: 
\begin{question}
Does the body $K$ from Question~\ref{question:KForSumOfBoundedOpnorm-has-LargeMeasure} 
always satisfy $w(K) \geq c\sqrt{n}$, where $c>0$ is a universal constant.
\end{question}

\bibliographystyle{alpha}
\bibliography{discrepancy-for-spectral}

\appendix

\section{Missing Proofs for Preliminaries\label{sec:MissingProofsOfPrelim}}

\begin{proof}[Proof of Cor.~\ref{cor:ProbGaussianHasToBeTruncated}]
Since $\E[\|\bm{y}\|_2] \le \E[\|\bm{y}\|_2^2]^{1/2} = \sqrt{m}$, we apply Theorem 4 to get, for $m \ge 7$,

\[
\Pr_{\bm{y} \sim N(\bm{0}, \bm{I}_m)} \Big[\|\bm{y}\|_2 > m\Big] \le e^{-(m-\sqrt{m})^2/2} \le 2^{-m}.
\]

Since the function $y \mapsto -\frac{y^2}{2} + \log(y^2)$ is concave, we can upper bound it with any tangent line; in particular, 

\[
-\frac{y^2}{2} + \log(y^2) \le \Big(\frac{2}{m} - m\Big) \cdot y+ \frac{m^2}{2}+ \log(m^2) - 2,
\]

so that using the standard estimate $P_{y \sim N(0,1)}[y > m] \ge \frac{m}{m^2+1} \cdot \frac{1}{\sqrt{2 \pi}} e^{-m^2/2}$, we have

\[
  \E_{y \sim N(0,1)} [y^2 \mid y > m] \le \frac{\displaystyle  \int_m^\infty \exp\Big(\Big(\frac{2}{m} - m\Big) \cdot y+ \frac{m^2}{2}+ \log(m^2) - 2 \Big) dy}{\sqrt{2\pi} P[y > m]} \le \frac{m^2+1}{m} \cdot \frac{m^3}{m^2-2}
\]
and therefore, for $m \ge 7$,
\[
\E[\|\bm{y}\|_2^2 \mid \|\bm{y}\|_2 > m] \le m \cdot \E_{y \sim N(0,1)} [y^2 \mid y > m] < 2m^3.
\]
Now, since

\[
m = \E\Big[\|\bm{y}\|_2^2\Big] = \underbrace{\Pr\Big[\|\bm{y}\|_2 > m\Big]}_{\le  \exp(-(m-\sqrt{m})^2/2)} \cdot \underbrace{\E\Big[\|\bm{y}\|_2^2 \mid \|\bm{y}\|_2 > m\Big]}_{\le 2m^3} + \underbrace{\Pr\Big[\|\bm{y}\|_2 \le m\Big]}_{\le 1}\cdot \E\Big[\|\bm{y}\|_2^2 \mid \|\bm{y}\|_2 \le m\Big],
\]
it follows that $\E\Big[\|\bm{y}\|_2^2 \mid \|\bm{y}\| \le m\Big] \ge (1 - 2^{-m})\cdot m$ for $m \ge 7$. 
\end{proof}


\end{document}